\documentclass[a4paper,12pt]{article} 

\usepackage{epsfig}

\usepackage{amssymb}
\begin{document}

\title{Search for Odderon induced contributions to \\ exclusive
Meson Photoproduction at HERA\footnote{Talk given
at the Intern. Conference on New Trends in High Energy Physics, \newline
22-29.9.2001, Yalta, Crimea, Ukraine.} }
\author{J. OLSSON \ \ (on behalf of the H1 collaboration) \\
   DESY, Notkestra\ss e 85, 22603 Hamburg, Germany \\
        E-mail: jan.olsson @ desy.de}
\date{ }
\maketitle 
\begin{abstract}
Odderon induced contributions to exclusive 
photoproduction of $\pi^{\circ}$, $f_2(1270)$ and $a_2(1320)$
have been searched for at HERA, using the multiphoton decays of these
mesons. No indication
for such contributions was found, in a kinematic region defined by the 
average photon-proton centre-of-mass energy
 $\langle W\rangle = 200 - 215$ GeV,
photon virtualities $Q^2 < 0.01\;$GeV$^2$ and   
$0.02\;$GeV$^2 < |t| < 0.3~\;$GeV$^2$, where $t$ is the 
squared momentum transfer at the proton vertex. The measured 
upper limits for the cross sections, 
$\sigma(\gamma p\to^{\hspace*{-2ex}\mathbb{O}}\ 
       \pi^{\circ}N^{\ast}) < 39$~nb,
$\sigma(\gamma p\to^{\hspace*{-2ex}\mathbb{O}}\ 
       f_2(1270)X) < 16$~nb and
$\sigma(\gamma p\to^{\hspace*{-2ex}\mathbb{O}}\ 
       a_2(1320)X) < 96$~nb,
 all at 95~\%~CL, 
are lower than the predictions
by a theoretical model.
Exclusive photoproduction of $\omega$ and $\omega\pi^{\circ}$, in the 
$3\gamma$ and $5\gamma$ decay modes, is observed
with the expected cross sections.  
\end{abstract}
%

\section{Introduction}
The discussion about the possible contribution of an odd-under-crossing
amplitude in high energy hadron-hadron scattering goes back to the early 
1970's. The seminal papers\cite{LUK73,JOY75} established the
Odderon\footnote{Odd-under-crossing-Pomeron\cite{JOY75}.} as the 
$C=P=-1$ partner of
the Pomeron trajectory, with an intercept $\alpha_{\mathbb{O}}(0) \approx 1$.
In the Regge picture the presence of the Odderon amplitude would lead
to a difference in the total cross sections for $hh$ and $\bar{h}h$ 
scattering at high energies, and thus to a violation of 
Pomeranchuk's theorem\cite{POM58}. This and other predictions based on the
differences in cross sections of $hh$ and $\bar{h}h$ scattering could however
not be satisfactorally tested, 
due to the scarcity of precise measurements at high
energies. Indeed, global fits (see e.g. \cite{DL92})
of the available $hh$ and $\bar{h}h$ data
seemed to establish the conventional Regge picture, with the high energy 
scattering dominated by exchange of the $C=P=+1$ Pomeron, and with 
the odd amplitudes  
dominated by (Reggeon) trajectories with intercepts 
$\alpha_{\mathbb{R}}(0) \approx 1/2$; Reggeon exchange then
contributes only at low scattering energies.
\par\noindent
In the parton picture, the quantum numbers of the Pomeron and Odderon
make it natural to view their exchange as exchange of 2 and 3 gluons, 
respectively.
With the development of perturbative QCD (pQCD)
the interest in the Odderon has in recent years intensified, 
since in the investigations of
multigluon compound states, exact solutions to the Odderon equations have 
been found\cite{LIP97}. pQCD based predictions, for
exclusive reactions specific
to Odderon exchange, like 
$ 
\sigma(\gamma p 
       \to^{\hspace*{-2ex}\mathbb{O}}\ \eta_{c}p)=50$~pb at $Q^2 = 0 
$\cite{BAR2001}, as well as for several asymmetry effects due to the 
interference of Pomeron and Odderon exchange\cite{INTERF}, 
now pose a challenge to experiments at HERA and elsewhere.
\medskip
\par\noindent
\begin{minipage}[t]{140cm}
\begin{minipage}[t]{5cm}
\epsfig{file=Odd.olsson1.eps,width=5cm}
\end{minipage}
\par\noindent
\hspace*{0.2cm}
\begin{minipage}[t]{5cm}
Figure 1: Diagram for exclusive $\pi^{\circ}$ photoproduction, via  
    Odderon-photon fusion. The proton is excited into an $I=1/2$ isobar.
\end{minipage}
\vspace*{-6cm}
\par\noindent
\hspace*{5.8cm}
\begin{minipage}[t]{8cm}
\par\noindent
Exclusive photoproduction and electroproduction of 
pseudoscalar and tensor mesons via Odderon-photon fusion are reactions where
Pomeron exchange 
cannot contribute, and their detection and measurement would therefore be a
clear proof of the existence of the Odderon.
Such reactions are accessible at the $ep$
collider HERA. The corresponding diagram is shown in Fig.~1, for the 
particular case of
exclusive $\pi^{\circ}$ photoproduction,
\begin{equation}
  \label{eq:process}
  ep \to^{\hspace*{-2ex}\mathbb{O}}\ e\pi^{\circ}N^{\ast}.
\end{equation}
\end{minipage}
\end{minipage}
\smallskip
\par\noindent
The cross section of process  (\ref{eq:process}) has been calculated by  
E.R. Berger et al.\cite{ERB99} and is of special interest to the HERA 
experiments, since the predicted cross section is sizable and within reach
of experimental confirmation. The calculation is based on non-perturbative
QCD, applying functional methods\cite{NAC91} in the framework of the 
"Model of the Stochastic Vacuum" (MSV)\cite{DOS88}. In the model the
proton is viewed as a diquark-quark system in transverse space; through 
symmetry arguments the suppression of Odderon exchange in the elastic case (as
well as in $pp$ and $\bar{p}p$ scattering) can be explained. 
This suppression is not present when the proton is excited into an 
$N^{\ast}$ state with negative
parity. Thus the predicted cross section is
$
\sigma (\gamma p 
       \to^{\hspace*{-2ex}\mathbb{O}}\ 
       \pi^{\circ}N^{\ast}) = 
  294 \ {\rm nb} 
$
at $W_{\gamma p} = 20$  GeV. Using the relation
$
\sigma = \sigma_{\circ} (W^2_{\gamma p}/20^2)^{\alpha_{\mathbb{O}}(0) - 1}, 
$
the same\footnote{In \cite{ERB99}  
$\alpha_{\mathbb{O}}(0)=1.15$ is estimated,
leading to an even larger predicted cross section  
at HERA energies. The use of $\alpha_{\mathbb{O}}(0)=1$ is however strongly
recommended\cite{DOS2001}.}  
large cross section is predicted for HERA energies, 
$\langle W_{\gamma p}\rangle = 200 - 215$  GeV.

\section{Monte Carlo Simulation}

A characteristic feature of the MSV calculation is the $t$ dependence of the
cross section, shown as $d\sigma/dp_T$ in Fig.~2 ($p_T$ is the transverse
momentum of the produced $\pi^{\circ}$, and $|t|\sim p^2_T$ is a good
approximation in photoproduction). As seen, the cross section is large in the
region of experimental acceptance, $0.02 < |t| < 0.3~\;$GeV$^2$, in contrast
to the case of the $\gamma\gamma$ reaction.
\vspace*{0.4cm}
\par\noindent
\begin{minipage}[t]{14cm}
\hspace*{8.3cm}
\begin{minipage}[t]{5.5cm}
 \epsfig{file=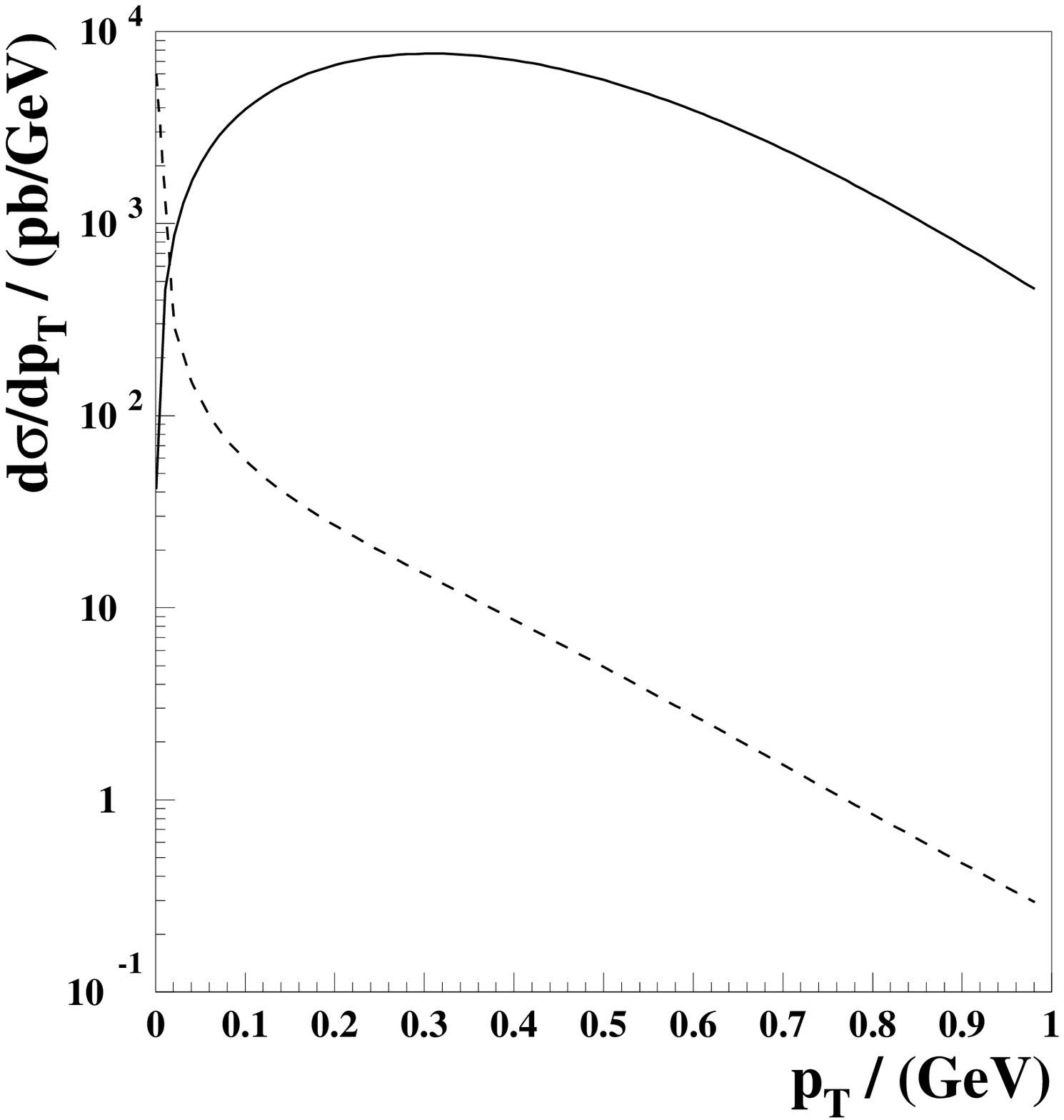,width=5.5cm}
\end{minipage}
\par\noindent
\hspace*{8.4cm}
\begin{minipage}[t]{5.4cm}
Figure 2: The $\pi^{\circ}$ transverse momentum distribution for 
Odderon exchange (solid line), compared to the corresponding distribution 
for photon exchange (dashed line).
\end{minipage}
\vspace*{-8.8cm}
\par\noindent
\begin{minipage}[t]{8cm}
\par\noindent
In order to simulate events of the Odderon exchange reaction (1), 
the Monte Carlo simulation
program DIFFVM\cite{LIS91} was modified to include this 
characteristic
$t$ dependence. Further modifications concern the inclusion of
 the $N^{\ast}$ states 
N(1520), N(1535), N(1650) and N(1700); 42\%\ of their decays result
in a leading neutron.
\par\noindent
Trivial background was simulated using the PYTHIA program\cite{SJO94}. Such 
background is expected from several diffractive processes with incompletely
reconstructed final states, like exclusive vector meson production 
($\gamma p \to \omega N^{\ast}, \ \omega \to \gamma \pi^{\circ}; \ 
 \gamma p \to \rho^{\circ} N^{\ast}, \ \rho^{\circ} \to \gamma \pi^{\circ}$)
or inclusive $\pi^{\circ}$ production, $\gamma p \to \pi^{\circ} X N^{\ast}$.
Exclusive $\pi^{\circ}$ production via $\gamma\gamma$ interactions (Primakoff
effect) or Reggeon ($\omega$) exchange, 
$
\gamma p \to^{\hspace*{-2ex}\omega} \ \pi^{\circ}N^{\ast} 
$,
is negligible.
\end{minipage}
\end{minipage}

\section{Experimental procedure}

A comprehensive description of the H1 detector is given in \cite{ABT97}.
The detector components of importance for the present analysis are given by
the simple experimental signature of process (\ref{eq:process}): The two 
photons from the $\pi^{\circ}$ decay are detected in the 
backward\footnote{The z-axis
of the H1 coordinate system coincides with the HERA beamline, 
the proton beam direction defining the positive direction.} 
electromagnetic calorimeters of H1, the SpaCal\cite{SPA96} and the 
VLQ\cite{VLQ98}, while the scattered electron is detected
in the Electron Tagger, located 33~m upstream of the interaction point.
The $N^{\ast}$ is identified through those decays in which a leading neutron
is produced. The latter is detected in the Forward Neutron 
Calorimeter (FNC), located 108~m downstream of the interaction point. 
The other major components of H1, namely the tracking chambers and the liquid 
argon calorimeter, were only used for the veto conditions, in the selection 
of exclusive events of process (\ref{eq:process}).
\par\noindent
The data used in this analysis correspond to an integrated luminosity of 
30.6~pb$^{-1}$ and were obtained during the data taking period 1999-2000.
The $e$ and $p$ beam energies were 27.5 and 920 GeV, respectively.  
The trigger 
was given by a combination of energy signals in the involved calorimeters,
namely VLQ, FNC and Electron tagger. 
For further details of the experimental setup and the event selection, 
see \cite{GOL2000}.
\par\noindent
\begin{minipage}[t]{14cm}
\begin{minipage}[t]{7cm}
\begin{minipage}[t]{7cm}
  \epsfig{file=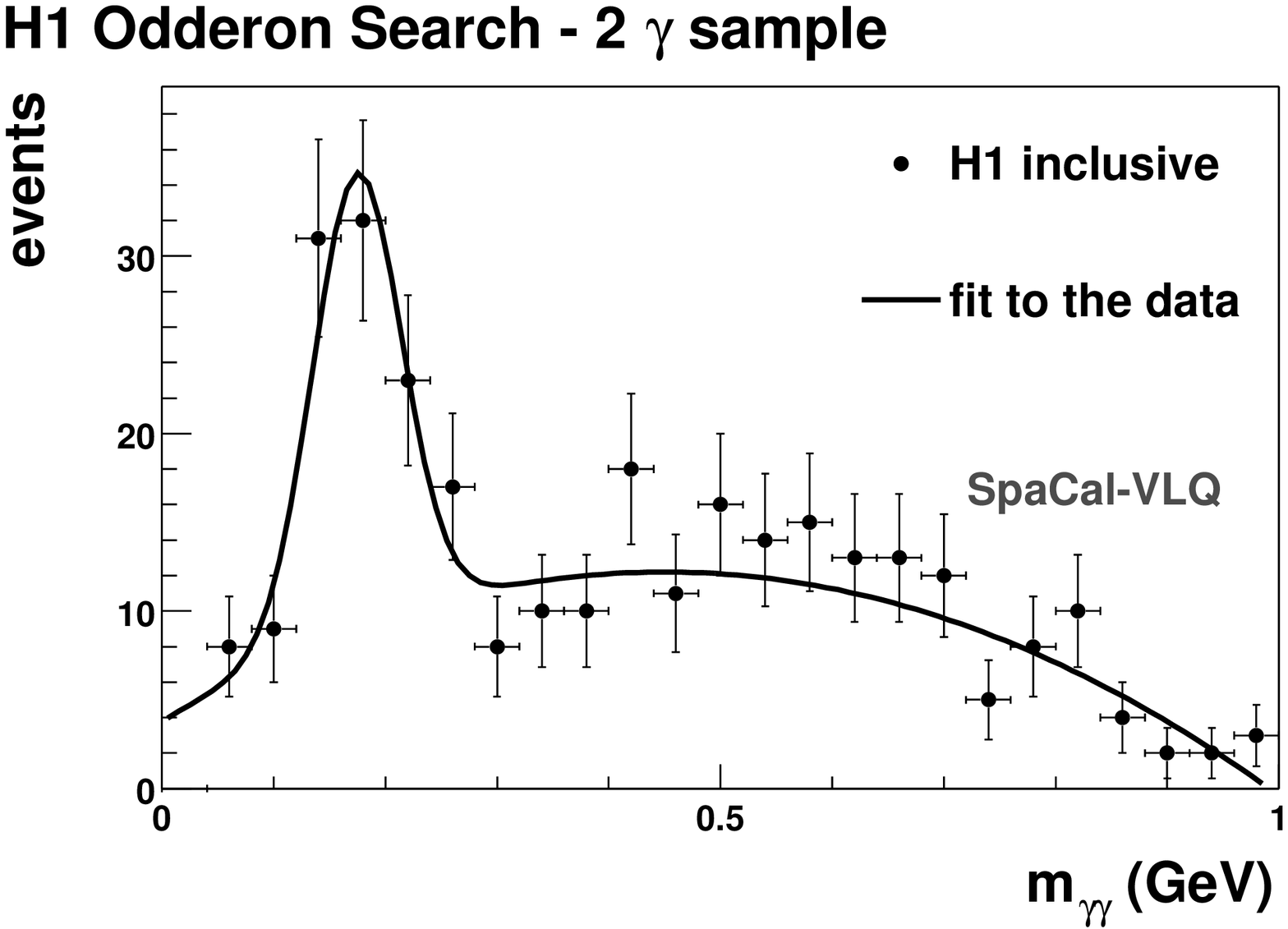,width=7cm}
\end{minipage}
\begin{minipage}[t]{7cm}
  \epsfig{file=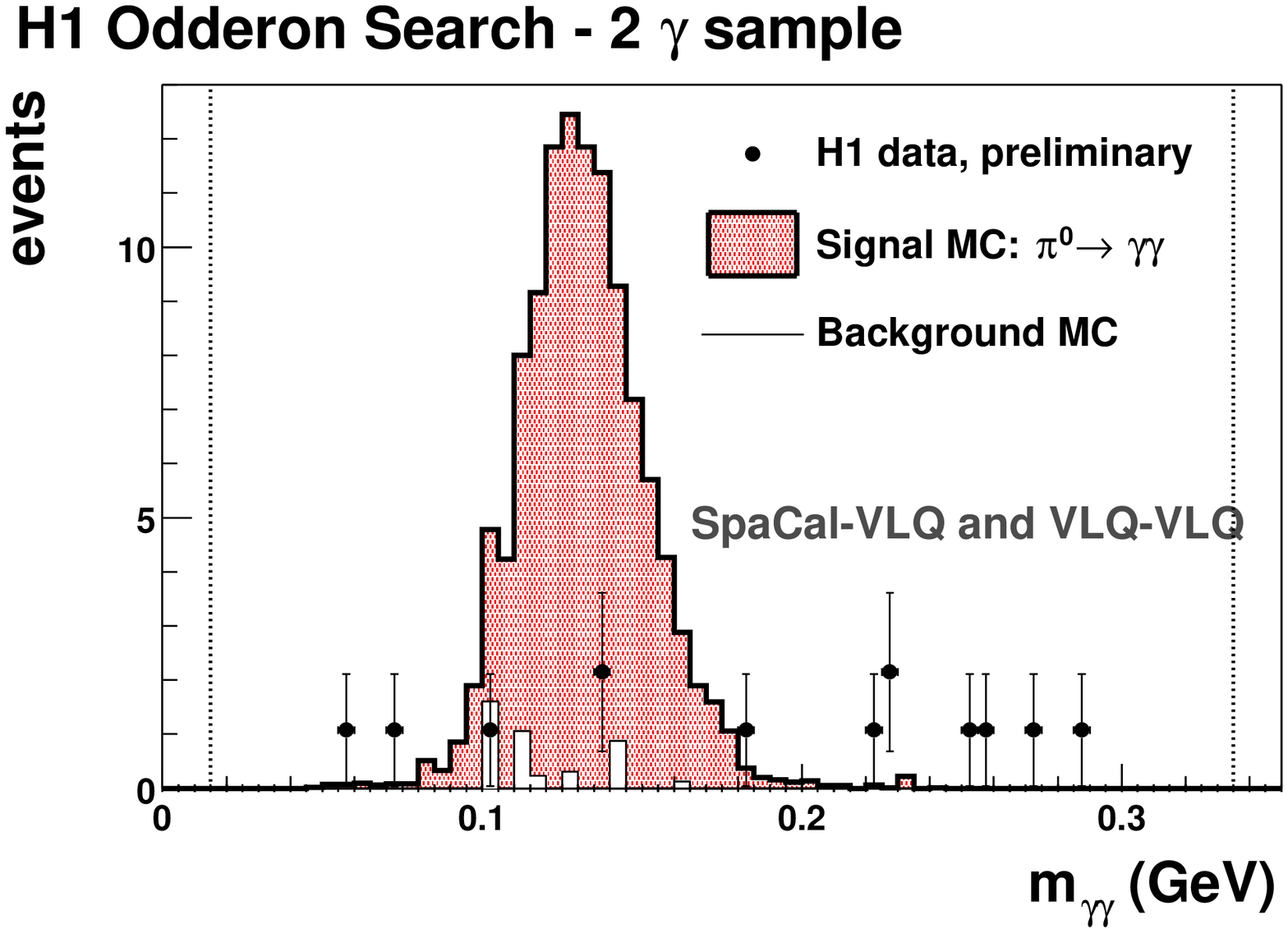,width=7cm}
\end{minipage}
\vspace*{-0.5cm}
\par\noindent
\hspace*{0.2cm}
\begin{minipage}[t]{6.4cm}
Figure 3: Two photon inv. mass. \newline
a) Inclusive two photon 
event sample with
one photon detected in VLQ and one in SpaCal. 
The curve shows the fitted sum of a
Gaussian and a polynomial background term. \newline
b) Final exclusive two photon 
event sample. One photon is
detected in VLQ, the other either in SpaCal or in VLQ. Also shown are the 
expectations from model (hatched histogram) and background (white
histogram). Vertical lines indicate the mass region for $\pi^{\circ}$
candidates.
\end{minipage}
\end{minipage}
\vspace*{-11cm}
\par\noindent
\hspace*{1.7cm}
{\large\bf b)}
\vspace*{-5.5cm}
\par\noindent
\hspace*{3cm}
{\large\bf a)}
\vspace*{-1.5cm}
\par\noindent
\hspace*{7.2cm}
\begin{minipage}[t]{6.7cm}
In the first step of the analysis events with exactly two "good"
photons are selected, where "good" photons are defined as electromagnetic
clusters with energy above certain thresholds, with cluster shapes compatible
with the photon hypothesis, and with positions within the fiducial volumes of
the calorimeters, avoiding energy loss from shower leakage. At least one
photon had to be detected in the VLQ (trigger condition). This selection
defines an {\it inclusive} event sample, for which the mass of the two photons
is shown in Fig. 3a, for the topology VLQ-SpaCal photons.  
\par\noindent
A clear $\pi^{\circ}$ signal is seen in Fig. 3a.
Since the events are not yet subjected
to an {\it exclusive} selection, this signal can be taken as proof that 
$\pi^{\circ}$'s indeed can be detected.
In order to arrive at a sample of exclusive events, corresponding to 
reaction (1), further selection cuts are applied: No charged track activity
is allowed in the event, and the longitudinal energy balance $E - p_z$ 
must be satisfied
by the energies of the scattered electron and the two photons:
$49~<~(E~-~p_z)_{e^{\prime}\gamma\gamma}~<~60$~GeV 
(the longitudinal energy balance of all particles
produced at the electron vertex is expected to sum up to twice the 
incident electron energy).
\end{minipage}
\end{minipage}
\vspace*{0.1cm}
\par\noindent
The invariant mass distribution of
the two photons in the final two
photon event sample, after 
all selection cuts, is shown if Fig. 3b.
Only a few events pass all cuts, and
there is no indication of a $\pi^{\circ}$ signal; altogether 13 events are 
observed in the generous $\pi^{\circ}$ window, indicated by the dotted lines.
These few events are consistent with the 
background expectation from PYTHIA, namely 4 events. 
In contrast, the expectation from the MSV model is 110 events.
\par\noindent
An upper limit for the cross section of reaction (1) can be derived from 
the data, using the prescription of \cite{COU98}. This preliminary 
upper limit, \newline
\centerline{$
\sigma (\gamma p 
       \to^{\hspace*{-2ex}\mathbb{O}} \ \pi^{\circ}N^{\ast}) < 39 \ {\rm nb} 
                \hspace*{0.5cm} (95 \%\ {\rm CL}),
$}
has to be compared with the predicted value of 200 nb at HERA energies;
the latter value is clearly incompatible with the observation 
even when considering the warning
given in \cite{ERB99}, ``uncertainty at least a factor 2''.

\section{Discussion}

The upper limit for the cross section of the reaction 
$
  ep \to^{\hspace*{-2ex}\mathbb{O}}\ e\pi^{\circ}N^{\ast}
$ 
is a factor 5 below the model prediction. Can this non-observation be
understood?  Possible explanations are:
\begin{enumerate}
\item
The energy dependence of the cross section is different from the assumed
one, implying that the Odderon intercept is considerably smaller than
unity. Indeed, predictions for the Odderon intercept in the literature 
span a wide range. The H1 non-observation result can also be interpreted as an
upper limit on  the intercept, $\alpha_{\mathbb{O}}(0) < 0.65$. Such a low
value is rather in the range of standard Reggeon intercepts, incompatible
with the expectation for the "classical" Odderon.
\item
The coupling $\gamma\mathbb{O}\pi^{\circ}$ is much smaller than assumed in 
\cite{ERB99}. This could be due to the Goldstone Boson nature of the     
$\pi^{\circ}$\cite{NAC2001}. More reliable predictions for Odderon induced
exclusive meson production can then be expected for heavier mesons, like the
tensor mesons $f_2(1270)$ and $a_2(1320)$. Predictions for such cross sections
have also been made within the MSV framework\cite{ERB2000,DOS2001} and their 
experimental investigation by the H1 collaboration\cite{BUD01,BER2001} 
is described below.    
\end{enumerate}

\section{Photoproduction of \boldmath$f_2(1270)$ and \boldmath$a_2(1320)$}

In order to study the exclusive photoproduction of the tensor mesons
$f_2(1270)$ and $a_2(1320)$, their decays into four photons were utilized:
$ep\to e f_2(1270)X$, $f_2\to 2\pi^{\circ}$ and 
$ep\to e a_2(1320)X$, $a_2\to \pi^{\circ}\eta$. The data set used in this 
analysis 
\vspace*{0.2cm}
\par\noindent
\begin{minipage}[t]{14cm}
\begin{minipage}[t]{7cm}
\epsfig{file=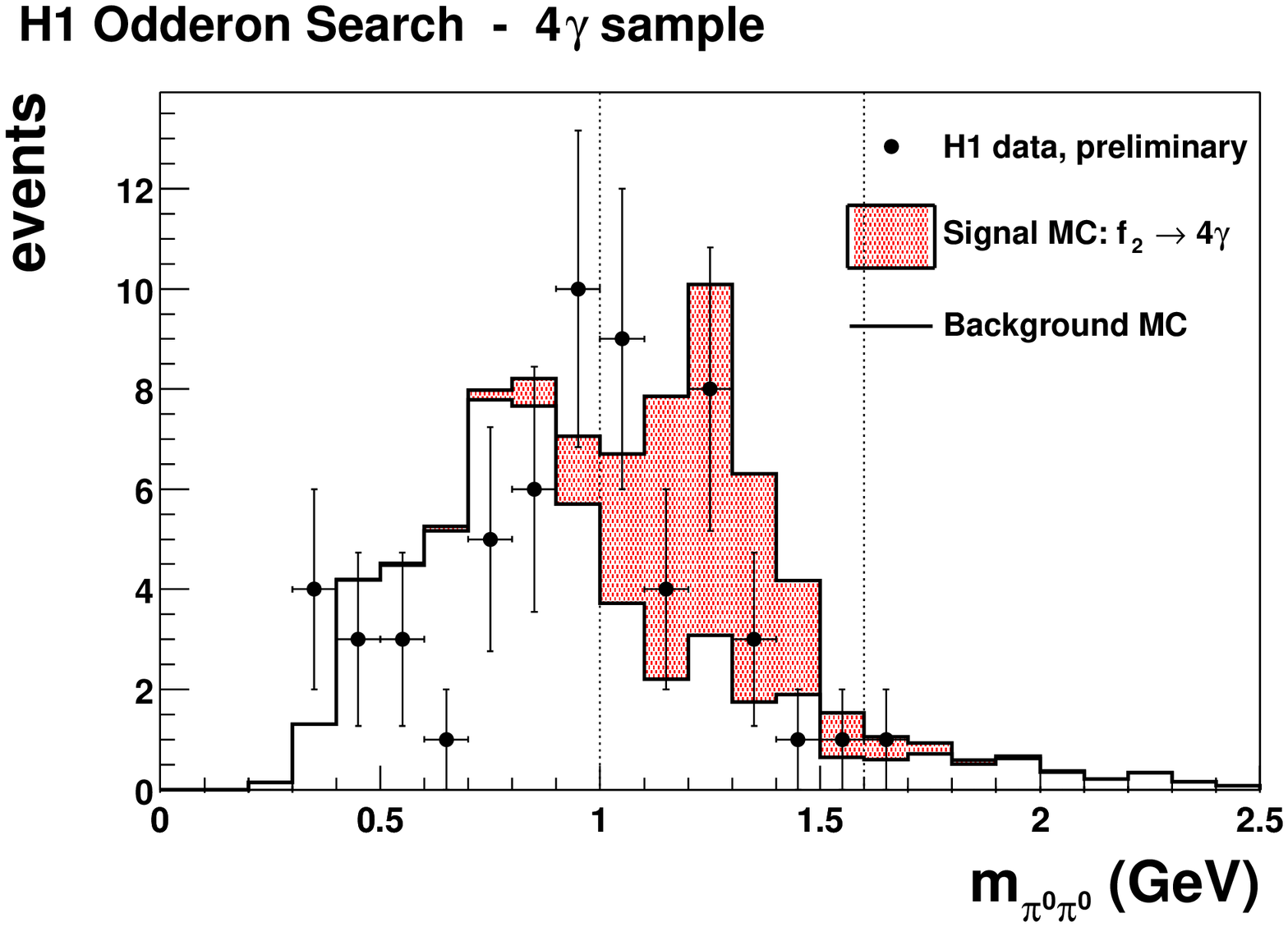,width=7cm}
\end{minipage}
\begin{minipage}[t]{7cm}
\epsfig{file=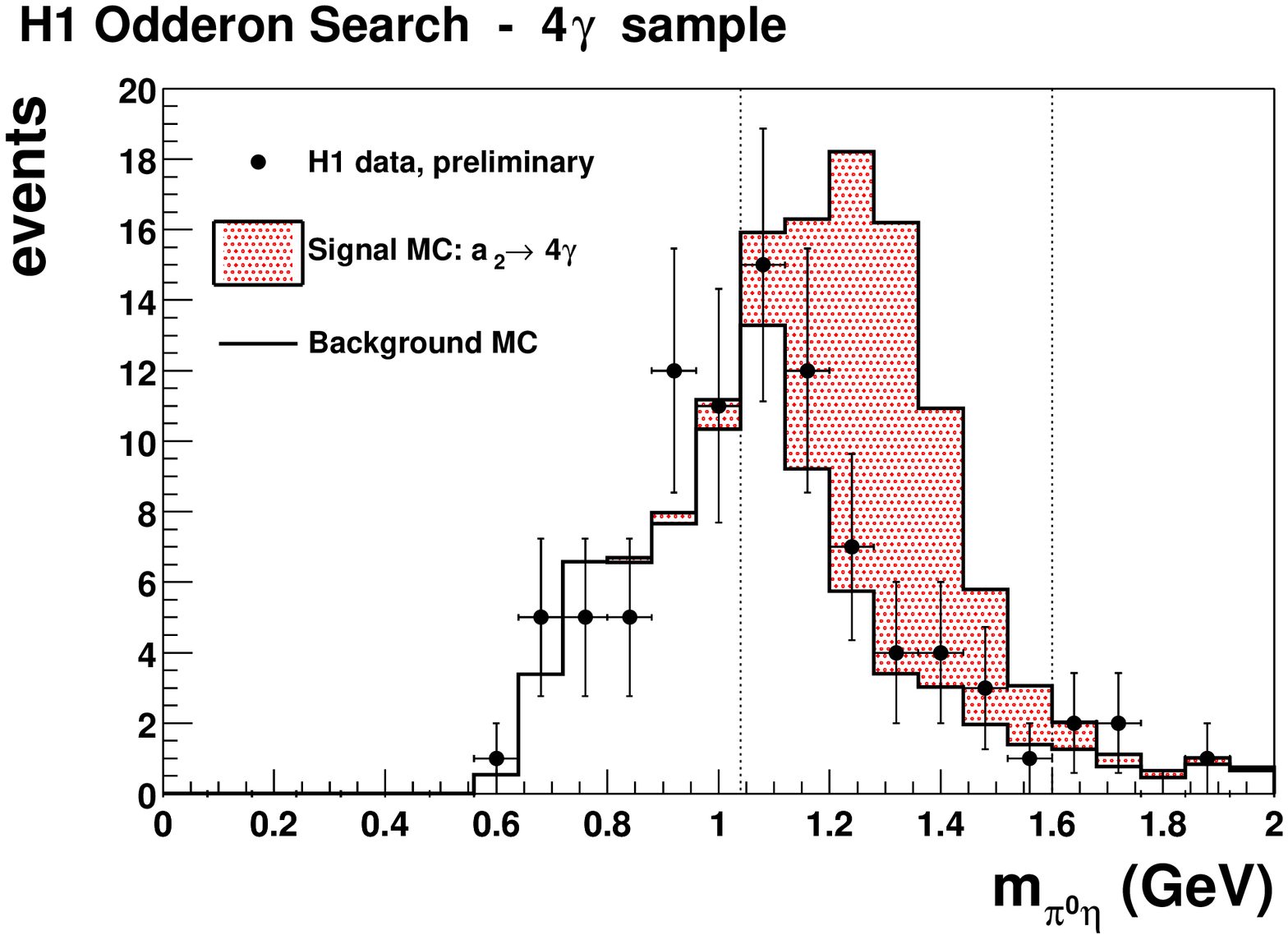,width=7cm}
\end{minipage}
\end{minipage}
\vspace*{-0.3cm}
\par\noindent
\hspace*{0.5cm}
\begin{minipage}[t]{13cm}
Figure 4: Four photon invariant mass, for 
$\pi^{\circ}\pi^{\circ}$  and
$\pi^{\circ}\eta$ event candidates.   
Also shown are the expectations from  
model (hatched histograms) and background (white
histograms). Vertical lines indicate the $f_2$ and $a_2$ mass bands, 
used for the upper limits
derivation.
\end{minipage}
\vspace*{0.3cm}
\par\noindent
was obtained in 1996 and corresponds to 4.5 pb$^{-1}$. In contrast
to the previous analysis, the four photons
were all detected in the SpaCal, and the trigger was based on a combination
of energy deposits in the electron tagger and the SpaCal, without special
requirements on a detected neutron. The average CMS energy is 
$\langle W_{\gamma p}\rangle = 200$ GeV, corresponding to the lower proton
beam energy in this data taking period, 820 GeV.
\par\noindent
Exclusive events with exactly four good photons were selected, satisfying the
restrictions of longitudinal energy balance and no charged track activity.
The two photon mass distribution, with 6 entries per event, has a clear
$\pi^{\circ}$ peak. However, the recoil mass spectrum, obtained for the
remaining two photons for each $\pi^{\circ}$ candidate, has only a weak
$\pi^{\circ}$ signal, and no $\eta$ signal. Selecting nevertheless 
those events with two photon
combinations in massbands corresponding to $\pi^{\circ}\pi^{\circ}$ and 
$\pi^{\circ}\eta$ final states, the four photon 
mass distributions in Fig.~4 are obtained.  
They agree well with the expected, trivial
background, as given by PYTHIA.
Since no signals of $f_2$ or $a_2$ are seen in Fig.~4,
preliminary upper limits for the cross sections are derived, 
again using \cite{COU98}:\newline
\centerline{
$\sigma(\gamma p\to^{\hspace*{-2ex}\mathbb{O}}\ 
       f_2(1270)X) < 16$~nb \ \ and \ \ 
$\sigma(\gamma p\to^{\hspace*{-2ex}\mathbb{O}}\ 
       a_2(1320)X) < 96$~nb.}
These limits,
both at 95~\%~CL, can be compared with the model 
predictions\cite{ERB2000,DOS2001}, 21 and 190 nb, respectively.

\section{Exclusive photoproduction of \boldmath$\omega$ and 
\boldmath$\omega\pi^{\circ}$}

Multiphoton final states induced by Odderon exchange have C-parity $+1$
and thus even number of photons. In contrast, final states induced by
Pomeron exchange have C-parity $-1$ and consequently an odd number of
photons. Exclusive final states of three and five photons were therefore
studied, searching for exclusive vector meson production, a process which
is conventionally described by Pomeron exchange. The 
same data set was used as in case of the four photon final state
investigation described above.
\vspace*{0.2cm}
\par\noindent
\begin{minipage}[t]{14cm}
\begin{minipage}[t]{7cm}
\begin{minipage}[t]{7cm}
\epsfig{file=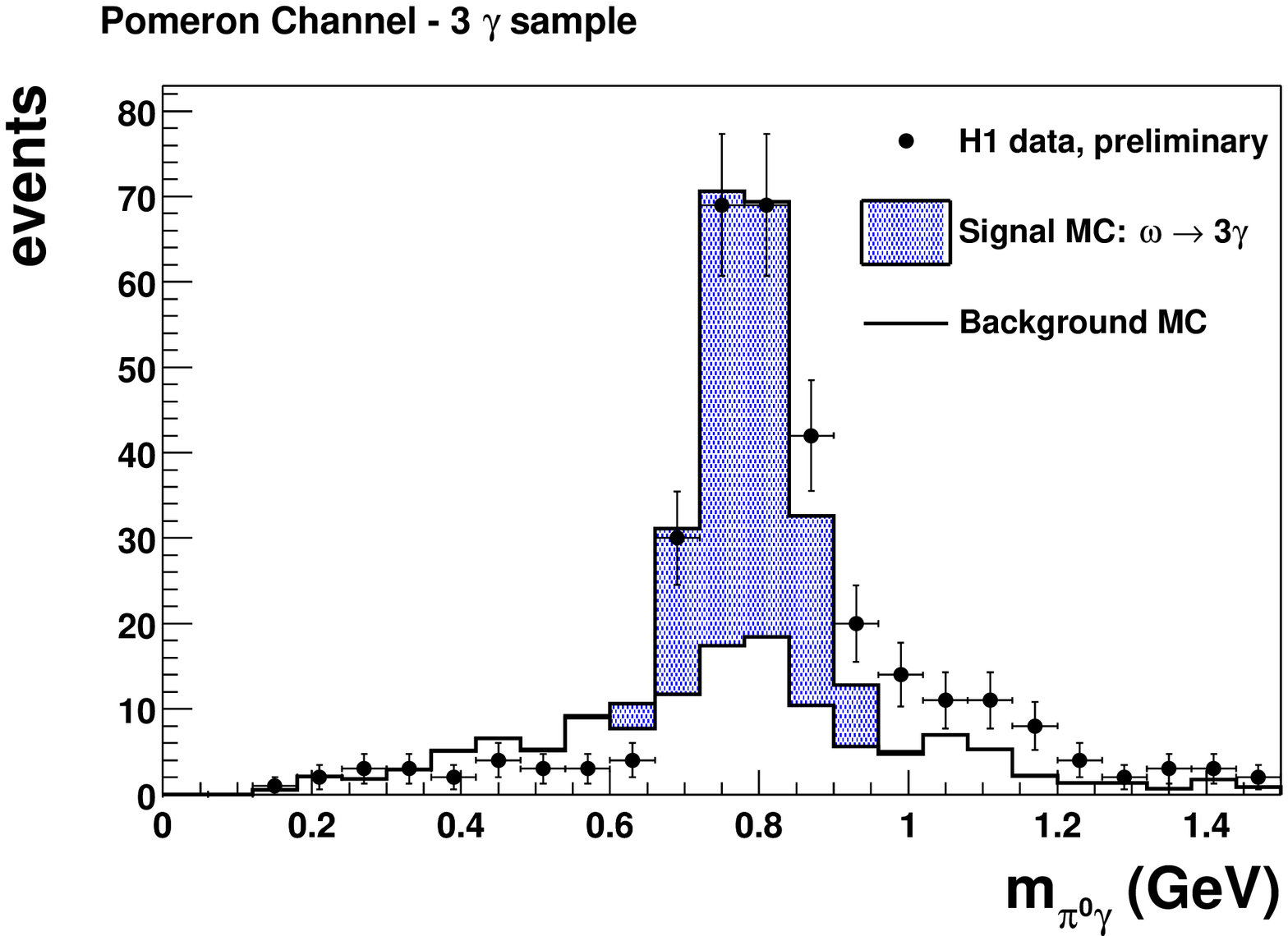,width=7cm}
\end{minipage}
\vspace*{-0.5cm}
\par\noindent
\hspace*{0.5cm}
\begin{minipage}[t]{6cm}
Figure 5: Three photon invariant mass, for  
$\gamma\pi^{\circ}$ event candidates. Model  
prediction and expected background (hatched and white
histograms) are also shown.
\end{minipage}
\end{minipage}
\vspace*{-7.6cm}
\par\noindent
\hspace*{7cm}
\begin{minipage}[t]{7cm}
\begin{minipage}[t]{6.8cm}
\epsfig{file=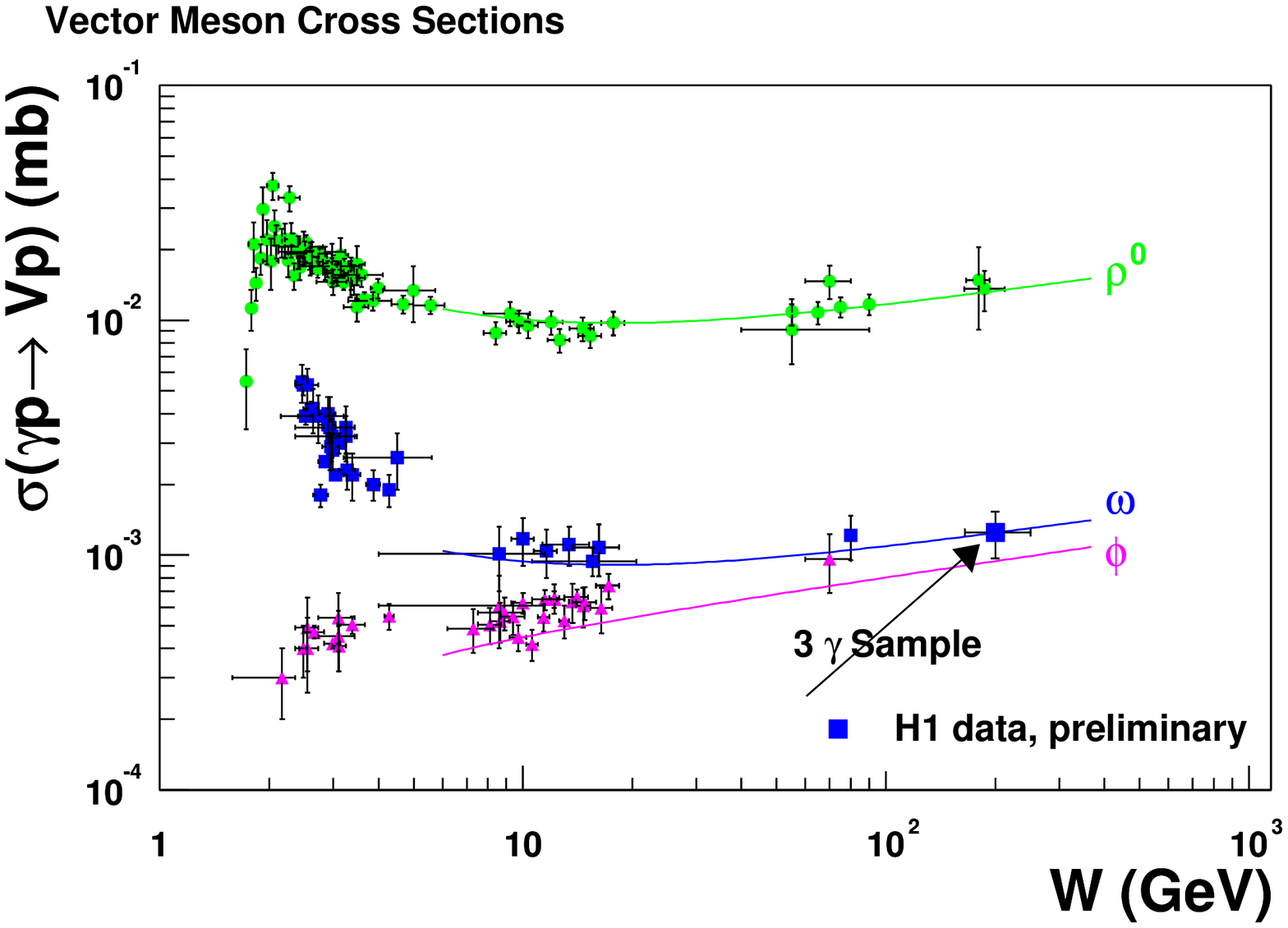,width=6.8cm}
\end{minipage}
\par\noindent
\hspace*{0.5cm}
\begin{minipage}[t]{6cm}
Figure 6: Synopsis of elastic vector meson photoproduction cross
sections vs. $W_{\gamma p}$, from fixed target to HERA experiments.
\end{minipage}
\end{minipage}
\end{minipage}
\vspace*{0.7cm}
\par\noindent
In the exclusive three photon event sample, 
the two photon mass distribution, with 3 entries per event, shows a clear
$\pi^{\circ}$ peak. The three photon mass distribution for events 
having at least one candidate for the $\gamma\pi^{\circ}$ final state
is shown in Fig.~5. 
A prominent $\omega$ peak is seen, due to the decay
$\omega \to \gamma\pi^{\circ}$.
The corresponding\footnote{Note that feed-down background from the
 exclusive photoproduction of
$\omega\pi^{\circ}\to 5\gamma$ has been taken into account.}
 preliminary cross section
is \newline
\centerline{
$\sigma(\gamma p \to \omega p) = 
                   (1.25 \pm 0.17 \pm 0.22) \ \mu{\rm b}$,}
which agrees very well with the expectation from other measurements,
Fig.~6.
\vspace*{0.05cm}
\par\noindent
\begin{minipage}[t]{14cm}
\begin{minipage}[t]{7cm}
\begin{minipage}[t]{7cm}
\epsfig{file=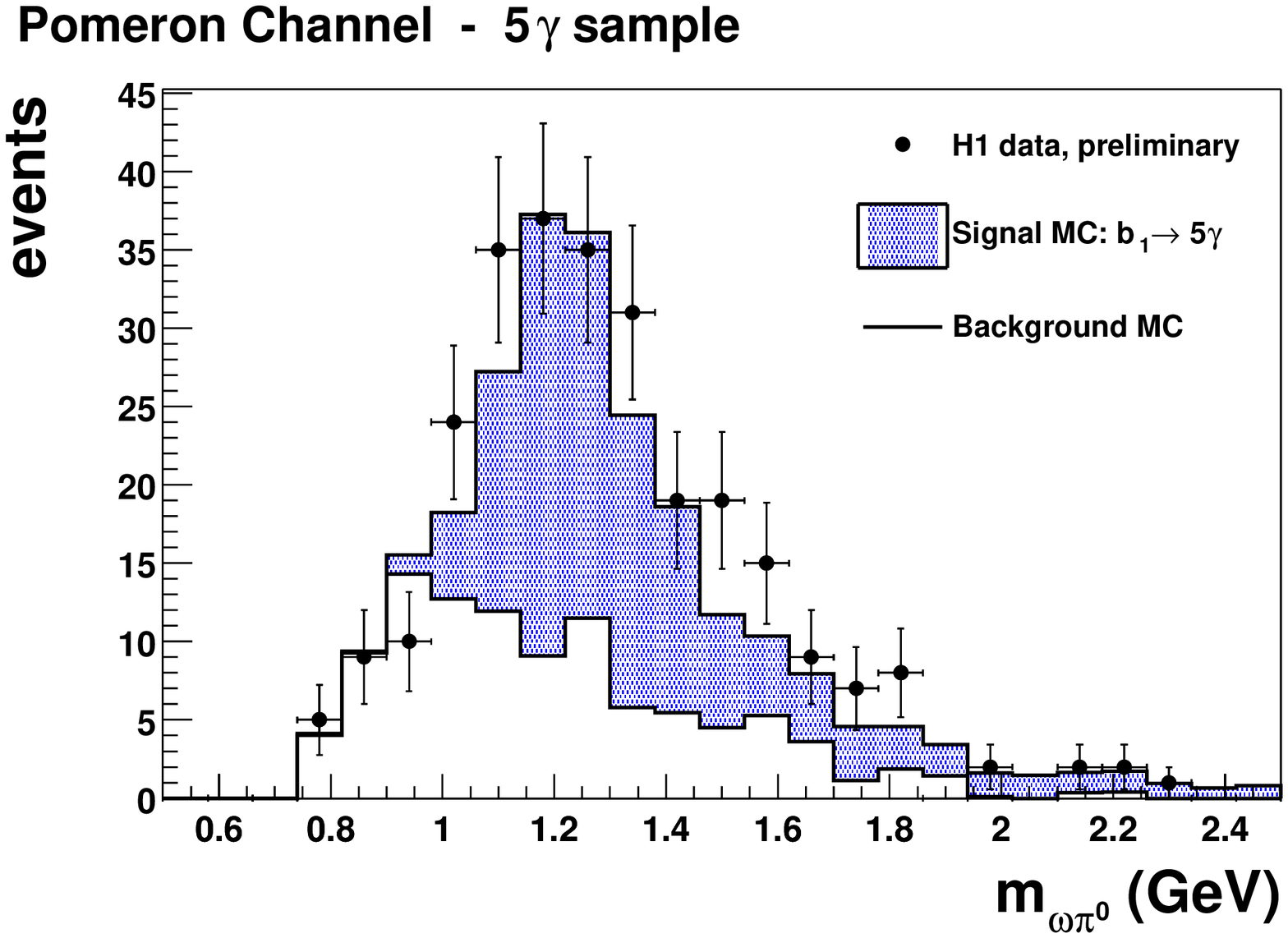,width=7cm}
\end{minipage}
\vspace*{-0.5cm}
\par\noindent
\hspace*{0.5cm}
\begin{minipage}[t]{6cm}
Figure 7: Five photon invariant mass for  
$\omega(\gamma\pi^{\circ})\pi^{\circ}$ event candidates.
Also shown are the model  
prediction (hatched histogram) and the expected background (white
histogram).
\end{minipage}
\end{minipage}
\vspace*{-8cm}
\par\noindent
\hspace*{7.2cm}
\begin{minipage}[t]{6.8cm}
Also the exclusive final state of five photons shows a prominent
$\pi^{\circ}$ peak in the two photon mass distribution (10 entries per
event). Many events are candidates for the final state 
$\gamma\pi^{\circ}\pi^{\circ}$ and the three photon mass  recoiling against
a $\pi^{\circ}$ candidate shows a clear $\omega$ peak. Selected
events with mass combinations in the corresponding $\pi^{\circ}$ and
$\omega$ mass bands have the five photon mass distribution in Fig.~7.
The axial vector meson $b_1(1235)\to \omega\pi^{\circ}$, earlier 
observed in photoproduction at lower energies\cite{OMEGA},
is a candidate for the broad resonant structure observed above the 
PYTHIA background.
\end{minipage}
\end{minipage}
\vspace*{0.4cm}
\par\noindent
The preliminary cross section is  
\newline
\centerline{$
\sigma(\gamma p \to \omega \pi^{\circ} X) = 
                   (980 \pm 200 \pm 200) \ {\rm  nb},
$}
with a background of 190 nb. The difference, i.e. the cross section which
possibly may be attributed to exclusive 
resonant $\omega\pi^{\circ}$ production,
is compatible with the 
cross section measured in \cite{OMEGA}
for $b_1(1235)$ photoproduction. Extrapolated
to HERA energies\footnote{The extrapolation followed \cite{CUD2000}
in analogy to the ordinary vector mesons with the $\gamma\mathbb{P}V$ coupling
adapted to match the low energy data.}, it is
   $\sigma(\gamma p \to b_1(1235) p) = (660 \pm 250)~{\rm nb}.$

\section{Summary}

A search for exclusive photoproduction of $\pi^{\circ}, f_2(1270)$ and 
$a_2(1320)$, induced by Odderon exchange, was performed using multi-photon
final states. No signal was 
found. Upper limits for the cross sections are below the predictions of a 
model based on non-perturbative QCD.
\par\noindent
Exclusive $\omega$ and $\omega\pi^{\circ}$ photoproduction is observed in 
three and five photon final states, at levels consistent with the 
expectations from Pomeron exchange and consistent with other observations of
exclusive vector and axial vector meson photoproduction.

\section{Acknowledgments}

It is a pleasure to thank the organizers for the warm and joyful 
atmosphere in a most interesting and remarkably well prepared conference.
Very useful information exchange with B.~Nicolescu, K.~Kang and
O.V. Teryaev is gratefully acknowledged. I also wish to thank my
colleagues in H1, for providing the data and the results shown in this report
and for all their help given to me.

\end{document}